\documentclass[aps,prd,reprint,preprintnumbers,groupedaddress,superscriptaddress,nofootinbib,amssymb,notitlepage,eqsecnum,bibnotes]{revtex4-2}

\pdfoutput=1

\usepackage[T1]{fontenc} 
\usepackage[dvipdfmx]{graphicx}
\usepackage{amsmath, amsthm, amsfonts, amssymb,amstext}
\usepackage{microtype}
\graphicspath{{./}}
\usepackage[colorlinks=true,citecolor=blue,urlcolor=blue,linkcolor=blue]{hyperref}
\usepackage{xcolor}
\usepackage{bm}
\allowdisplaybreaks[1]

\newcommand{\mpl}{M_{\rm Pl}}
\newcommand{\Lag}{\mathcal{L}}
\newcommand{\mS}{\mathcal{S}}
\newcommand{\mK}{\mathcal{K}}
\newcommand{\dd}{\mathrm{d}}

\begin{document}

\preprint{WUCG-26-08}

\title{Isotropic Universes with a Preferred Direction}

\author{Jose Beltr\'an Jim\'enez}
\email{jose.beltran@usal.es}
\affiliation{Departamento de F\'isica Fundamental and IUFFyM, Universidad de Salamanca, E-37008 Salamanca, Spain}

\author{Shinji Tsujikawa}
\email{tsujikawa@waseda.jp}
\affiliation{Department of Physics, Waseda University, 3-4-1 Okubo, Shinjuku, Tokyo 169-8555, Japan}

\begin{abstract}
We present a class of cosmological scenarios in which a preferred spatial direction in the matter sector coexists with an exactly homogeneous and isotropic Friedmann--Lema\^{\i}tre--Robertson--Walker (FLRW) geometry. The Cosmological Principle is realized on shell, with the vector-field equations enforcing a vanishing momentum density and suitable interactions eliminating the anisotropic stress. 
Consequently, the construction yields an entire FLRW branch without tuning the matter-field initial conditions. The preferred direction reemerges in perturbations, producing direction-dependent propagation and mixing among scalar, vector, and tensor modes already at linear order. In particular, the mixing opens a linear channel through which perturbations in the scalar sector can in principle source gravitational waves. Our construction thus reveals a new route by which preferred-direction physics can leave observable cosmological signatures while remaining hidden in the background geometry.
\end{abstract}

\maketitle

%%%%%%%%%%%%%
\section{Introduction}
%%%%%%%%%%%%%

Physical systems often hide microscopic complexity at large distances, where simpler and more symmetric structures emerge. The emergence of symmetry  
is a powerful organizing principle across many areas of physics and guides the construction of effective field theories (EFTs). In cosmology, it underlies the Cosmological Principle, according to which the Universe is homogeneous and isotropic on sufficiently large scales. For spatially flat cosmologies, this leads to a FLRW geometry with Euclidean ISO(3) symmetry.

Since matter sources gravity through its energy-momentum tensor, it is sufficient at the background level for this tensor, rather than the underlying fields themselves, to respect the symmetries required by the Cosmological Principle. Homogeneous scalars provide the standard, trivial realization of this principle. More generally, composite operators such as the energy-momentum tensor can possess more symmetry than their constituents, allowing field configurations to break spatial rotations and/or translations while preserving diagonal combinations with internal transformations \cite{Nicolis:2015sra}.

The realization of these background symmetries can nevertheless have important consequences for perturbations. For example, vector triads and SU(2) gauge-field backgrounds can preserve spatial isotropy in a nontrivial way through a diagonal combination of spatial rotations and internal rotations or gauge transformations \cite{Cervero:1978db,Galtsov:1991un,Darian:1996mb,Armendariz-Picon:2004say,Golovnev:2008cf,Maleknejad:2011jw,Maleknejad:2011sq,Maleknejad:2012fw,Adshead:2013qp,Mehrabi:2015lfa,Nieto:2016gnp,Alvarez:2019ues,Garcia-Serna:2025dhk}. 
A characteristic consequence of this diagonal realization of rotational symmetry is the emergence of an additional tensor mode, even though the theory contains no extra fundamental spin-2 field. Its mixing with the standard gravitational-wave (GW) mode gives rise to tensor oscillations and related propagation effects \cite{Maleknejad:2011sq,Maleknejad:2012fw,Adshead:2013qp,Caldwell:2016sut,BeltranJimenez:2019xxx,Ezquiaga:2021ler,Aoki:2025uwz,deCesare:2025ovv}. Solid \cite{Endlich:2012pz,Bucher:1998mh,Gruzinov:2004ty,BeltranJimenez:2026tlr,Esposito:2026diu} and gaugid \cite{Piazza:2017bsd} cosmologies extend this logic to nontrivial realizations of translations and display distinctive perturbation signatures, with related constructions also occurring in teleparallel cosmology \cite{Gomes:2023hyk}. Some of these realisations can be related via dualities \cite{Aoki:2022ylc}.

All these constructions realize cosmological symmetries off shell, namely: once the field configuration and the relevant internal symmetries are specified, the diagonal symmetry holds exactly without using the background equations of motion. However, another possibility is to realize isotropy only approximately, as it occurs for rapidly oscillating vector or higher-spin fields, whose energy-momentum tensor becomes isotropic after averaging over oscillations with a time scale much shorter than the Hubble time \cite{Cembranos:2012kk,Cembranos:2012ng,Cembranos:2013cba,Cembranos:2016ugq}. 
These scenarios show that background isotropy need not imply a manifestly isotropic matter sector, but their isotropy is approximate and relies on the field dynamics.

In this Letter, we identify a sharper possibility. A matter sector with a preferred direction can support an exact FLRW geometry for which the Cosmological Principle is realized only on shell. Isotropy is neither an off-shell symmetry of the fields nor an averaged property, but a dynamical consequence of the field equations. The mechanism does not select a special matter-field trajectory, since every physical matter solution on the homogeneous FLRW branch has an energy-momentum tensor compatible with an exact FLRW geometry, even though a generic off-shell field configuration is anisotropic. The preferred direction remains latent in the background but reemerges in linear perturbations, producing direction-dependent propagation and scalar--vector--tensor helicity mixing. Our construction thus separates two notions that are usually tied together: an exactly isotropic background metric and an isotropic perturbation theory with helicity decoupling.

%%%%%%%%%%%%%%%%%%%%%%%%%%%%%%%%
\section{An on-shell realization of the Cosmological Principle}
%%%%%%%%%%%%%%%%%%%%%%%%%%%%%%%%

Let us consider a triplet of vector fields $A^a{}_{\mu}$ with $a=1,2,3$, endowed with an internal SO(3) symmetry, in the following homogeneous configuration:
\begin{equation}
A^a{}_{\mu}=\phi^a(t)\delta^0_{\mu}+A(t)\delta^a_{\mu}\,.
\label{genconfig}
\end{equation}
On the $A(t)=0$ branch, the vector fields have only homogeneous temporal components, which transform as scalars under spatial rotations, so isotropy is realized trivially. If $\phi^a(t)=0$, we recover the standard triad with isotropy realized as a combination of spatial and internal rotations. When both $A(t)$ and $\phi^a(t)$ are present, the internal rotation required to compensate a spatial rotation of the triad also rotates $\phi^a$, so full isotropy is broken. At each time, $\phi^a(t)$ defines a genuine preferred direction with an instantaneous SO(2) little group. The field strength $F^a{}_{\mu\nu} =\partial_{\mu}A^a{}_{\nu}-\partial_{\nu}A^a{}_{\mu}$ satisfies $F^a{}_{ij}=0$ and $F^a{}_{0i}=\dot A\delta^a_i$, where an overdot denotes time-derivative, so it is invariant under the diagonal rotations of the triad and independent of $\phi^a$. We use $\delta^a{}_i$ to identify spatial and internal indices, so that $\phi^i\equiv\delta^i{}_a\phi^a$ and $\phi_i\equiv\delta_i{}^a\phi_a$.

For the configuration \eqref{genconfig} to source an FLRW metric, 
its energy-momentum tensor must have neither momentum density 
nor anisotropic stress. The first requirement follows from a general diffeomorphism Bianchi identity established in Ref.~\cite{Orjuela-Quintana:2024qfn} for homogeneous configurations that can be written as
\begin{equation}
\frac{\delta\mS}{\delta A^a{}_0}A^a{}_i
+2\frac{\delta \mS}{\delta g_{0\nu}}g_{i\nu}\equiv0,
\end{equation}
for any diffeomorphism-invariant vector-field action $\mS[A^a{}_\mu,g_{\mu\nu}]$, with $a$ running over internal indices. Using the definition $T_{\mu\nu}=-\tfrac{2}{\sqrt{-g}} \tfrac{\delta\mS}{\delta g^{\mu\nu}}$, this expression can equivalently be written as the off-shell identity
\begin{equation}
T^0{}_i\equiv-\frac{1}{\sqrt{-g}} \frac{\delta\mS}{\delta A^a{}_0}A^a{}_i\,.
\label{eq:T0iconstraint}
\end{equation}
Upon imposing the temporal vector equations,
$\tfrac{\delta\mS}{\delta A^a{}_0}=0$, the off-shell identity
\eqref{eq:T0iconstraint} becomes the on-shell relation $T^0{}_i=0$.
Hence, for homogeneous matter configurations on the FLRW branch, the vector-sector momentum density
vanishes on shell independently of the detailed form of the action. This
dynamically removes one potential obstruction to an FLRW geometry.

To ensure full isotropy, the anisotropic stress must also vanish
without selecting a special time evolution. Although this is not automatic,
we now show that suitable interactions can cancel it for arbitrary
homogeneous matter configurations on the FLRW branch without constraining
the background evolution.
To illustrate the mechanism, we consider the following representative two-derivative EFT Lagrangian:
\begin{equation}
\Lag=-\frac14\sum_{i=0}^{4}a_iY_i -V(X_1,X_2)\,,
\label{lag}
\end{equation}
where
\begin{equation}
\begin{aligned}
Y_0&=F^a{}_{\mu\nu}F_a{}^{\mu\nu},
&Y_1&=A^a{}_{\alpha}A_a{}^{\alpha}
      F^b{}_{\mu\nu}F_b{}^{\mu\nu},\\
Y_2&=A^a{}_{\mu}A_a{}^{\nu}F^b{}^{\mu\alpha}F_{b\nu\alpha},
&Y_3&=A^a{}_{\alpha}A^b{}^{\alpha}
      F_{a\mu\nu}F_b{}^{\mu\nu},\\
Y_4&=A^a{}_{\alpha}A^b{}_{\beta}
      F_a{}^{\alpha\nu}F_b{}^{\beta}{}_{\nu},\\
X_1&=A^a{}_{\alpha}A_a{}^{\alpha},
&X_2&=A^a{}_{\alpha}A_{a\beta}
      A^{b\alpha}A_b{}^{\beta}\,.
\end{aligned}
\label{invariants}
\end{equation}
Here, the $a_i$ are constant parameters, and $V$ is a function of
$X_1$ and $X_2$. 
More general functions of these operators, higher-order terms, and 
nonminimal curvature couplings can be included, but the Lagrangian \eqref{lag} 
already displays the mechanism. 
We consider the spatially flat FLRW metric
\begin{equation}
\dd s^2=-\dd t^2+a^2(t)\dd\vec{x}^{\,2}\,,
\label{flrw}
\end{equation}
where $a(t)$ is the time-dependent scale factor. The spatial components of the energy-momentum tensor have the form $T^i{}_j=p \delta^i{}_j+\Pi\phi^i\phi_j$, where $p$ is the isotropic contribution. The anisotropy coefficient $\Pi$ can be read off from the off-diagonal components, which are
\begin{equation}
T^i{}_{j}=\frac{\phi^i\phi_j}{2a^2}
\left[(2a_3+a_4)\dot A^{2}-8A^2V_{,X_2}\right],
\qquad i\neq j\,,
\label{anisstress}
\end{equation}
where $V_{,X_2}=\partial V/\partial X_2$.
This contribution trivially vanishes on either the $\phi^a(t)=0$ branch
or the $A(t)=0$ branch, since the configuration \eqref{genconfig} is isotropic in those cases. For arbitrary $\phi^a\neq0$ and $A\neq0$, the coefficient of $\phi^i\phi_j$ in Eq.~\eqref{anisstress} vanishes under the conditions
\begin{equation}
2a_3+a_4=0\qquad\text{and}\qquad V_{,X_2}=0\,.
\label{conditions} 
\end{equation}
These relations constrain the interactions rather than the background
solution. They cancel the full anisotropic stress for arbitrary
homogeneous matter evolution on the FLRW branch and hence do not select a
special cosmological trajectory. Once
they hold, $p$ coincides with the physical isotropic pressure $P$ used below.
On the FLRW background and under these conditions, the momentum density and the $\phi^a$ equation of motion are
\begin{align}
T^0{}_{i}&=\frac{A\phi_i}{2a^2}\left[3(2a_1+a_2)\dot A^2-4a^2V_{,X_1}\right],
\label{momentum}\\
0&=\phi^a\left[3(2a_1+a_2)\dot A^2-4a^2V_{,X_1}\right]\,.
\label{phiEOM}
\end{align}
The common factor in square brackets makes the proportionality between
$T^0{}_i$ and the $\phi^a$ equation of motion manifest, in agreement with
Eq.~\eqref{eq:T0iconstraint}. Together with the cancellation of the
anisotropic stress, this establishes our central result: the theory
\eqref{lag} subject to the conditions \eqref{conditions} has a homogeneous
and isotropic energy-momentum tensor for every homogeneous matter
solution on this branch, even though the underlying vector
configuration contains a preferred direction.
The construction therefore yields an entire FLRW branch rather than an
isolated isotropic trajectory selected by initial conditions.
Although we have used \eqref{lag} as a representative example of the on-shell realization of the Cosmological Principle, the construction applies to more general Lagrangians.

To show that our construction does not result in a trivial cosmology, we note that, besides the trivial branch $\phi^a=0$, Eq.~\eqref{phiEOM} admits a nontrivial branch satisfying
\begin{equation}
V_{,X_1}=\frac{3(2a_1+a_2)}{4a^2}\dot A^2 .
\label{branch}
\end{equation}
Since $X_1=3A^2/a^2-\delta_{ab}\phi^a\phi^b$,
Eq.~\eqref{branch} involves only the magnitude of $\phi^a$ and leaves its
orientation undetermined. We write the internal vector as
\begin{equation}
\phi^a=\phi\,\hat n^a,\qquad
\phi\equiv(\delta_{ab}\phi^a\phi^b)^{1/2},\qquad
\delta_{ab}\hat n^a\hat n^b=1\,.
\label{hatn}
\end{equation}
For a generic potential that allows Eq.~\eqref{branch} to be inverted, the
constraint determines $\phi^2$ implicitly as a function of $A$, $\dot A$,
and $a$, whereas the orientation $\hat n^a(t)$ remains undetermined. We denote
the corresponding comoving spatial unit vector
$\hat n^i\equiv\delta^i{}_a\hat n^a$ by $\bm{\hat n}(t)$.
This freedom gives rise to an emergent on-shell invariance of the background
equations under $\phi^a(t)\to R^a{}_b(t)\phi^b(t)$, where
$R^a{}_b(t)$ represents a time-dependent SO(3) rotation. Unlike the internal
SO(3) symmetry of the action, which would also act on $A^a{}_i$, the resulting
invariance acts only on $\phi^a(t)$. 
The preferred direction $\hat n^a(t)$ is
therefore selected spontaneously in each background solution while remaining
invisible to the FLRW metric and its background evolution. Perturbations can
nevertheless distinguish different orientations, as we will show below. The same degeneracy may also
permit cosmological domains with different preferred directions, within each
of which the background symmetries are realized on shell.

Finally, imposing the conditions \eqref{conditions} and using the nontrivial-branch relation \eqref{branch}, the Lagrangian \eqref{lag} yields the following energy density and pressure of the vector sector:
\begin{align}
\rho&=
V+\left[\mu+\frac{a_2(a^2X_1-2A^2)}{a^2}-\frac{4a_3A^2}{a^2}\right]
\frac{3\dot A^2}{4a^2},\label{rho}\\
P&=-V+\left[\mu+\frac{a_2(a^2X_1+2A^2)}{a^2}+\frac{4a_3A^2}{a^2}\right]
\frac{\dot A^2}{4a^2},\label{press}
\end{align}
where $\mu\equiv 2(a_0+a_1X_1)$. Once the constraint is used, they depend 
only on rotational scalars, consistently with the exact FLRW geometry. 
As a minimal illustration, consider
\begin{equation}
V=V_0+b_1X_1^2\,,
\label{Vexample}
\end{equation}
where $V_0$ and $b_1$ are constants.
The constraint \eqref{branch} gives the relation 
$\phi^2=3[8b_1A^2-(2a_1+a_2)\dot A^2]/(8b_1a^2)$, so
Eqs.~\eqref{rho}-\eqref{press} reduce to
\begin{align}
\rho&=V_0+\rho_{A_0}+\rho_{A_1}+\rho_{A_2},\label{rhocon}\\
P&=-V_0+\frac{\rho_{A_0}}{3}-\frac{\rho_{A_1}}{3}
-\frac{\rho_{A_2}}{9},\label{presscon}
\end{align}
where
\begin{align}
\rho_{A_0}&=\frac{3a_0\dot A^2}{2a^2},\qquad
\rho_{A_1}=-\frac{3(a_2+2a_3)A^2\dot A^2}{2a^4},
\notag\\
\rho_{A_2}&=\frac{27(2a_1+a_2)^2\dot A^4}{64b_1a^4}\,.
\label{rhoAi}
\end{align}
The contributions $V_0$, $\rho_{A_0}$, $\rho_{A_1}$, and $\rho_{A_2}$
enter $P$ with the respective coefficients $-1$, $1/3$, $-1/3$, and
$-1/9$, as seen from Eqs.~\eqref{rhocon}--\eqref{rhoAi}. These coefficients
do not by themselves define separately conserved components. In
Appendix~\ref{Appendix}, we discuss the dark-energy dynamics at the background
level and show that the time-dependent vector sector can remain negligible
during radiation domination, while the fractional density associated with
$\rho_{A_1}$ grows during matter domination and $V_0$ eventually drives
cosmic acceleration. This demonstrates that the model is not merely a formal
realization of on-shell isotropy but can also support a nontrivial late-time
cosmological history.

\section{A minisuperspace approach}

Our on-shell realization of the Cosmological Principle admits a transparent
minisuperspace interpretation. Since only homogeneity is imposed off shell,
we describe the metric by the homogeneous ADM lapse, shift, and spatial
metric, $N(t)$, $N_i(t)$, and $q_{ij}(t)$, respectively
\cite{Arnowitt:1962hi}, and denote the reduced matter Lagrangian by
$\Lag_{\rm mini}$. Provided derivatives enter only through
$F^a{}_{\mu\nu}$, its dependence on the matter variables is through $A(t)$,
$\dot A(t)$, and $\phi^a(t)$, but not through $\dot\phi^a(t)$.

The $\phi^a$-dependent invariants that can source anisotropic stress may be chosen as
$q_1=q^{ij}\phi_i\phi_j$ and
$q_2=q^{ik}\delta_{k\ell}q^{\ell j}\phi_i\phi_j/2$.\footnote{A dependence on $q_2$ requires higher-order operators, such as $Z\equiv F^{a\mu\rho}F^{a\nu\sigma}F^b{}_{\mu\lambda}F^{c\lambda}{}_\nu A^b{}_\rho A^c{}_\sigma$. For example, if $\Lag=\mK(X_i,\bar{Y}_j,\bar{Z})$, where $\bar{Y}_j\equiv Y_j/Y_0$ and $\bar{Z}\equiv Z/Y_0^2$, the vanishing of the anisotropic stress on the FLRW background requires $\mK_{,X_2}=0$ and $6\mK_{,\bar{Y}_3}+3\mK_{,\bar{Y}_4}-\mK_{,\bar{Z}}=0$.}
Higher powers of the matrix $q^{ik}\delta_{kj}$ can be reduced by
the Cayley--Hamilton theorem. The metric variations of
$\phi^a$-independent invariants are proportional to $\delta_{ij}$ on the
FLRW background and hence contribute only to the isotropic pressure.
Variation with respect to the shift yields the momentum-density condition,
which gives no independent constraint after the temporal vector equations
are imposed, as follows from Eq.~\eqref{eq:T0iconstraint}. On the FLRW
background, $q_1=a^{-2}\phi^2$ and $q_2=a^{-4}\phi^2/2$, so the off-diagonal
spatial components take the form
\begin{equation}
\left.T_{ij}\right|_{\rm FLRW}\propto
\left(\frac{\partial\Lag_{\rm mini}}{\partial q_1}
+\frac{1}{a^2}\frac{\partial\Lag_{\rm mini}}{\partial q_2}\right)
\phi_i\phi_j,
\qquad i\neq j\,.
\label{mini}
\end{equation}
Together with the temporal vector equations, an FLRW solution
requires the coefficient in Eq.~\eqref{mini} to vanish on shell. For the
class of theories described by \eqref{lag}, we have
$\partial\Lag_{\rm mini}/\partial q_2=0$, while requiring
$\partial\Lag_{\rm mini}/\partial q_1=0$ for arbitrary homogeneous matter
evolution on the FLRW branch is equivalent to the two conditions in
Eq.~\eqref{conditions}.
Equation~\eqref{mini} therefore provides a minisuperspace criterion
for constructing EFTs of on-shell FLRW phases with latent directions.
Together with the temporal vector equations, it identifies when the
momentum density and anisotropic stress vanish on shell, while the
perturbation action remains sensitive to the anisotropic operators encoding
the hidden direction.

Unlike the off-shell and averaged realizations discussed in the Introduction,
our branch yields an exact FLRW background only on shell. Identifying a
symmetry principle that selects the corresponding EFT operators is an
important direction for future work.

%%%%%%%%%%%%%%%%%%%%%%%%%%
\section{Anisotropic propagation and helicity mixing}
%%%%%%%%%%%%%%%%%%%%%%%%%%

Let us now show how the latent direction emerges at the perturbative level. 
In standard FLRW perturbation theory, invariance of the full background configuration under spatial rotations 
prevents linear couplings among scalar, vector, and tensor perturbations, so the three sectors evolve independently.
In our scenario, the background metric remains rotationally invariant, whereas the
preferred direction $\hat n^a$ defined in Eq.~\eqref{hatn} becomes
observable through perturbations. The quadratic Lagrangian governing linear perturbations can be schematically decomposed as
\begin{equation}
\Lag_2=\Lag_{\rm iso}+\Lag_{\vec\phi}+\Lag_{\rm mix}\,.
\label{Ldecomp}
\end{equation}
The first term contains contributions that are independent of $\hat n^a$. 
It includes effects such as GW oscillations involving an additional tensor mode \cite{BeltranJimenez:2019xxx,Ezquiaga:2021ler,deCesare:2025ovv}. Such oscillations are already present in off-shell triad cosmologies \cite{Maleknejad:2011sq,Maleknejad:2012fw,Adshead:2013qp,Caldwell:2016sut} and are not, by themselves, a distinctive signature of the on-shell branch, so we do not discuss them further here. The remaining two terms encode the dependence on the preferred direction. Their distinctive feature in our scenario is that they coexist with an exactly FLRW background. The term $\Lag_{\vec\phi}$ is diagonal in helicity space but makes the propagation coefficients direction dependent, as also occurs in anisotropic models whose background contains an explicit preferred direction. By contrast, $\Lag_{\rm mix}$ couples different helicity sectors and thereby breaks the usual scalar--vector--tensor decoupling at linear order even though the background metric is exactly FLRW.

To illustrate the different pieces in Eq.~\eqref{Ldecomp} more explicitly, we focus on the sector involving tensor modes. For each Fourier mode with
comoving wave vector $\bm{k}$, we introduce a 
right-handed orthonormal frame
$\{\bm{\hat e}_1,\bm{\hat e}_2,\bm{\hat k}\}$, where
$k\equiv|\bm{k}|$ and $\bm{\hat k}\equiv\bm{k}/k$ is chosen as the $z$ axis.
In this frame,
\begin{equation}
\bm{\hat n}(t)=\sin\theta\left(
\cos\vartheta\,\bm{\hat e}_1+\sin\vartheta\,\bm{\hat e}_2\right)
+\cos\theta\,\bm{\hat k}\,.
\end{equation}
Here $\theta=\theta(t)$ and $\vartheta=\vartheta(t)$ are the polar and
azimuthal angles of $\bm{\hat n}(t)$, respectively.
We denote the spatial components of the vector-field and metric perturbations
by $\delta A^a{}_j$ and $\delta g_{ij}$, respectively, and define
$\delta A_{ij}\equiv\delta_{ai}\delta A^a{}_j$. We use the normalizations
$(\delta g_{ij}/a^2)^{\rm TT}=\sum_\lambda
h_{(\lambda)}e_{ij}^{(\lambda)}$ and
$(\delta A_{ij})^{\rm TT}=\sum_\lambda
t_{(\lambda)}e_{ij}^{(\lambda)}$, where the superscript ${\rm TT}$ denotes
the transverse--traceless part and the polarization tensors are normalized as
$e_{ij}^{(\lambda)}e^{ij(\lambda')}
=2\delta_{\lambda\lambda'}$. We take
$e_{ij}^{(+)}=\hat e_{1i}\hat e_{1j}-\hat e_{2i}\hat e_{2j}$ and
$e_{ij}^{(\times)}=\hat e_{1i}\hat e_{2j}+\hat e_{2i}\hat e_{1j}$. We use the Fourier convention
$f(t,\bm{x})=\int \tfrac{\dd^3k}{(2\pi)^3}\,f(t,\bm{k})
e^{-i\bm{k}\cdot\bm{x}}$, so that $\partial_i\to-ik_i$.
All Fourier arguments below are suppressed and 
understood to be $\bm{k}$.
An overbar denotes complex conjugation, and 
reality in position space implies
$\bar f(t,\bm{k})=f(t,-\bm{k})$ for any real perturbation $f(t,\bm{x})$. The expressions below are understood as
contributions to the quadratic action integrated over all $\bm{k}$. The
$\bm{k}$ and $-\bm{k}$ contributions are therefore complex conjugates, so no
explicit complex-conjugate term is added.
For the representative theory \eqref{lag}, after imposing
\eqref{conditions} and evaluating the quadratic action on the nontrivial
FLRW branch, the tensor contribution to the preferred-direction sector
$\Lag_{\vec\phi}$ is
\begin{align}
\Lag_{\vec\phi}^{\rm tensor}={}&
\sum_{\lambda=+,\times}\bigg[
\frac{ik\phi\cos\theta}{a}\,a_2A\bar t_{(\lambda)}
\dot t_{(\lambda)}
\notag\\
&-\frac{ik\phi\cos\theta}{a}(a_2+2a_3)A\dot A\,
\bar t_{(\lambda)}h_{(\lambda)}
\notag\\
&.-\frac{a_3\phi^2k^2\cos^2\theta}{2a}
t_{(\lambda)}\bar t_{(\lambda)}\bigg].
\label{tensorlag}
\end{align}
The displayed preferred-direction terms vanish for propagation perpendicular to the preferred direction, and their magnitudes are maximal for propagation along it. Hence, the theory predicts anisotropic propagation even though the background expansion is exactly isotropic.

The helicity-mixing sector is even more distinctive. We parametrize the antisymmetric helicity-0 component of $\delta A_{ij}$ as $(\delta A_{ij})_{\rm scalar}=\epsilon_{ijk}\partial^k S$, where $\partial^k\equiv\delta^{k\ell}\partial_\ell$, $\epsilon_{ijk}$ is the three-dimensional Levi--Civita symbol with
$\epsilon_{123}=+1$, and $S$ is the corresponding helicity-0 amplitude.
With the same Fourier convention, the resulting tensor--scalar mixing
term is
\begin{equation}
\Lag^{\text{t-s}}_{\rm mix}=ik^3\frac{a_3\phi^2\sin^2\theta}{a}
\Big[\cos(2\vartheta)t_{(\times)}
-\sin(2\vartheta)t_{(+)}\Big]\bar S\,.
\label{mixlag}
\end{equation}
This mixing is maximal for modes propagating orthogonally to $\bm{\hat n}(t)$ and vanishes for propagation along it, yielding an angular
dependence complementary to that in Eq.~\eqref{tensorlag}. The angle
$\vartheta$ determines the linear combination of tensor polarizations that
mixes with the scalar mode. We take $\bm{\hat e}_1$ and $\bm{\hat e}_2$ to be
time independent. For $\sin\theta\neq0$, a constant rotation of these vectors around $\bm{\hat k}$ can set $\vartheta=0$ at a reference time, so that only
$t_{(\times)}$ appears in Eq.~\eqref{mixlag} at that time. 
If $\vartheta$ subsequently evolves, maintaining this condition would require a co-rotating
transverse basis, whose time dependence would generate additional terms in the quadratic action. Thus, the time-dependent orientation of $\bm{\hat n}(t)$ cannot be eliminated by a choice of basis.

Equation~\eqref{mixlag} directly displays linear mixing between
the scalar-type vector-field perturbation $S$ and the tensor modes
$t_{(\lambda)}$. Since the latter also mix with $h_{(\lambda)}$ in the tensor
sector, this opens a linear channel through which scalar-sector perturbations
can in principle source metric GWs. The mechanism can generate
scalar--tensor correlations at linear order, whereas GWs induced by scalar
perturbations in standard FLRW cosmology arise only at second order.

The terms $\Lag_{\vec\phi}$ and $\Lag_{\rm mix}$ therefore lead to
complementary observational effects. The former produces direction-dependent
propagation, with coefficients that depend on the angle between the wave
vector and the preferred direction, whereas the latter allows perturbations
to convert among scalar, vector, and tensor sectors. Both effects coexist
with an exactly FLRW geometry because the background metric is insensitive
to $\bm{\hat n}(t)$, whereas perturbations are not.

Their coexistence provides a clean phenomenological handle. If the on-shell
sector is negligible in the early Universe and becomes dynamically relevant
only at late times, as illustrated at the background level by the dark-energy
application in Appendix~\ref{Appendix}, primordial perturbations can be
generated as usual and later undergo direction-dependent propagation and
helicity conversion. If $S$ is excited through its coupling to the
conventional scalar sector, primordial scalar perturbations constrained by
the cosmic microwave background may seed a tensor component correlated with
the matter distribution and carrying a characteristic angular dependence.
Conversely, GWs emitted by astrophysical sources can convert into additional
modes as they propagate through the cosmological background. The observed
signal may then exhibit not only changes in propagation speed or damping but
also changes in its mode and polarization content controlled by the line of
sight relative to $\bm{\hat n}(t)$. The same theory parameters govern the
preferred-direction effects across the scalar and tensor sectors, so a
combined analysis of growth, lensing, and GW propagation can test the
mechanism without relying on background anisotropy. Such signatures are
absent in a homogeneous and isotropic perfect-fluid description.

Related perturbative effects of cosmological vector backgrounds, including
direction-dependent GW propagation, polarization conversion, and
scalar--vector--tensor mixing, have been studied in
Refs.~\cite{Cembranos:2016ugq,Miravet:2020kuj,Miravet:2022pli,Chase:2023puj,Chase:2024wsq,Chase:2026qxs}. The distinctive point here is not these perturbative phenomena by themselves but their background realization. The treatments in those works range from time-averaged FLRW descriptions to Bianchi~I geometries with shear, whereas our construction supports direction-dependent propagation and helicity mixing on an exact FLRW branch without time averaging or background shear.

%\vspace{0.4cm}
%%%%%%%%%%%%
\section{Discussion}
%%%%%%%%%%%%
%\vspace{-0.2cm}

The central message of this Letter is simple: the Cosmological Principle
constrains what the background geometry sees, while perturbations can reveal
more. An exactly homogeneous and isotropic background geometry can coexist with a latent preferred direction whose structure is
revealed only by perturbations.

Several extensions naturally suggest themselves. Because the
symmetry is realized only on shell, a systematic EFT description will likely
require a formulation distinct from standard cosmological EFTs
\cite{Cheung:2007st,Gubitosi:2012hu,Tsujikawa:2014mba,Finelli:2018upr,Salcedo:2024smn}.
One possible route is to organize the operators on homogeneous but
anisotropic backgrounds (see, e.g.,
\cite{Abolhasani:2015cve,Rostami:2017wiy,Gong:2019hwj}) and then use the
momentum constraint together with Eq.~\eqref{mini} to identify on-shell FLRW
phases. This may enable a systematic classification of cosmologies with
latent directions. It would also be interesting to investigate whether
analogous mechanisms operate in braneworld and condensed-matter systems or
in compact objects whose spherical symmetry emerges only on shell. More
immediately, the full coupled perturbation system should be analyzed on
realistic backgrounds to establish the stable parameter space and construct
transfer functions for comparison with next-generation cosmological and GW
data. These developments would open a new avenue for exploring systems in
which spacetime symmetries emerge only on shell while perturbations reveal
their hidden structure.

\vspace{-0.3cm}
\begin{acknowledgments}
\vspace{-0.3cm}

We thank Antonio L. Maroto for useful discussions. J.B.J. acknowledges support from grants PID2024-158938NB-I00 funded by MICIU/AEI/10.13039/ 501100011033 and by “ERDF A way of making Europe” and Project SA097P24. S.T. acknowledges support from JSPS KAKENHI Grant Nos.~26K07090 and 26H00847, and from the Waseda University Special Research 
Projects (No.~2026C-486).
\end{acknowledgments}

\bibliography{PreferredRefs}

\appendix

%%%%%%%%%%%%%%%%%%%
\section{An application to dark energy}
\label{Appendix}
%%%%%%%%%%%%%%%%%%%

To illustrate how the on-shell sector can be embedded in a late-time
cosmology, we couple it to Einstein gravity and standard matter,
\begin{equation}
\mS=\int \dd^4x\sqrt{-g}\left(\frac{\mpl^2}{2}R
+\Lag\right)+\mS_{M}\,,
\label{action}
\end{equation}
where $g\equiv\det(g_{\mu\nu})$, $\mpl$ is the reduced Planck mass, and 
$R$ is the Ricci scalar.
The matter action $\mS_{M}$ contains separately conserved pressureless matter
and radiation, whose energy densities are denoted by $\rho_m$ and $\rho_r$,
respectively.
The background equations are
\begin{align}
3\mpl^2H^2&=\rho+\rho_m+\rho_r,\label{fried1}\\
\mpl^2(2\dot H+3H^2)&=-P-\frac{\rho_r}{3},\label{fried2}\\
\dot\rho+3H(\rho+P)&=0\,, 
\label{cont}
\end{align}
where $H=\dot{a}/a$, and $\rho$ and $P$ denote the energy density and pressure
of the vector sector. Although the underlying vector configuration contains
a hidden direction, its background evolution is therefore formally
indistinguishable from that of an isotropic dark sector.

%%%%%%%%%%%%%%%%%%%%%%%%%%%%%
\begin{figure}[ht!]
\centering
\includegraphics[height=3.0in,width=3.4in]{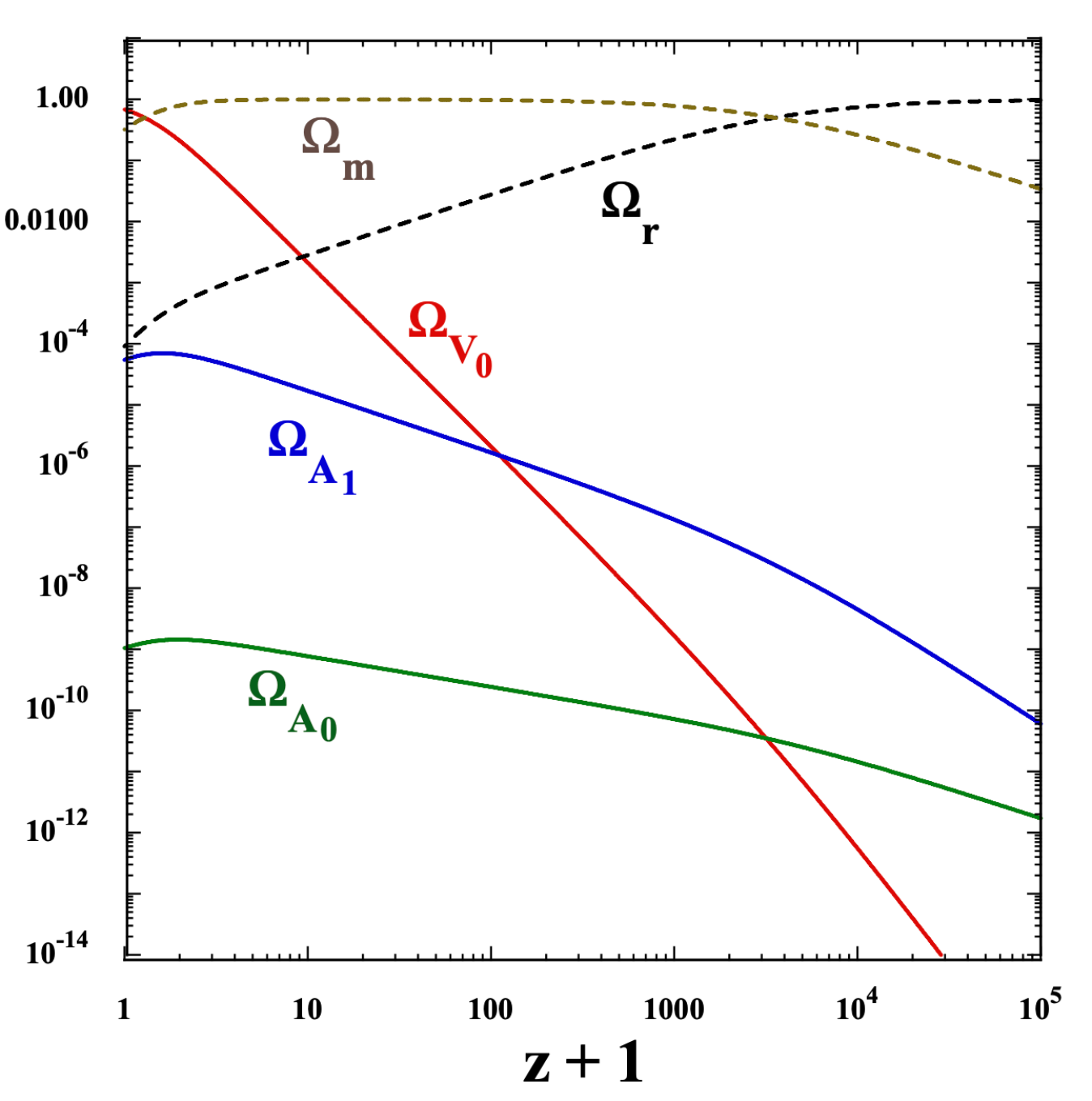}
\caption{Evolution of the density parameters
$\Omega_{V_0}$, $\Omega_{A_0}$, $\Omega_{A_1}$, 
$\Omega_m$, and $\Omega_r$ as functions of $1+z$ for a representative
solution of the model \eqref{Vexample}, where $z$ is the redshift.
The approximate present-day values are
$\Omega_{V_0}\simeq0.68$, $\Omega_m\simeq0.32$, 
$\Omega_r\simeq9.1\times10^{-5}$, $\Omega_{A_0}\simeq10^{-9}$, 
and $\Omega_{A_1}\simeq5.5\times10^{-6}$. 
The strongly suppressed $\Omega_{A_2}$ is not shown.}
\label{fig:densities}
\end{figure}
%%%%%%%%%%%%%%%%%%%%%%%%%%%%%%%%

We now specialize to the model \eqref{Vexample}. Consider a regime in which
$\rho_{A_1}$ dominates the time-dependent part of the vector energy density
and the contributions of $\rho_{A_0}$ and $\rho_{A_2}$ to
Eq.~\eqref{cont} are negligible. The constant term $V_0$ contributes $-V_0$
to $P$ and therefore cancels identically from the continuity equation, so it
need not be subdominant to $\rho_{A_1}$. Since $\rho_{A_1}$ contributes
$-\rho_{A_1}/3$ to $P$, Eq.~\eqref{cont} then reduces to
$\dot\rho_{A_1}+2H\rho_{A_1}\simeq0$.
Using Eq.~\eqref{rhoAi}, this relation is equivalent to
$\dd(A\dot A/a)/\dd t\simeq0$ and hence gives
$A^2=c_0+c_1\int a\,\dd t$, where $c_0$ and $c_1$ are integration constants.
For a power-law background $a\propto t^p$, the growing solution scales as
$A\propto t^{(p+1)/2}$. Defining
$\Omega_{A_i}\equiv\rho_{A_i}/(3\mpl^2H^2)$,
$\Omega_{V_0}\equiv V_0/(3\mpl^2H^2)$,
$\Omega_m\equiv\rho_m/(3\mpl^2H^2)$, and
$\Omega_r\equiv\rho_r/(3\mpl^2H^2)$, we obtain
$\Omega_{A_0}\propto t^{1-p}$,
$\Omega_{A_1}\propto t^{2(1-p)}$,
$\Omega_{A_2}\propto t^{-2p}$, and
$\Omega_{V_0}\propto t^2$.

During matter domination, $p=2/3$, so that
$\Omega_{A_1}\propto a$. This scaling remains applicable even when $V_0$
exceeds $\rho_{A_1}$ within the vector sector, provided that the expansion is
still matter dominated and $\rho_{A_0}$ and $\rho_{A_2}$ remain negligible.
It ceases to apply once the expansion departs appreciably from the
matter-dominated power law. The scaling of $\Omega_{A_0}$ also illustrates
why it need not follow $\Omega_r$. Although the $\rho_{A_0}$ term enters $P$
with the coefficient $1/3$, it does not obey a separate continuity equation.
The quantities $\rho_{A_i}$ are algebraic contributions to the energy density
of the same vector field and are not separately conserved. Consequently,
$\rho_{A_0}$ need not scale as $a^{-4}$, and $\Omega_{A_0}$ need not track
the radiation density parameter.  

The numerical solution in Fig.~\ref{fig:densities} illustrates these
behaviors. The nonconstant vector contributions remain subdominant in the
early Universe, while $\Omega_{A_1}$ grows during matter domination. Although
$V_0$ becomes the largest contribution within the vector sector well before
the onset of acceleration, this does not invalidate the matter-era scaling
derived above. At late times, $V_0$ drives cosmic acceleration and
$\Omega_{A_1}$ decreases as the background departs from matter domination.
The model can thus exhibit dark-energy dynamics with a time-dependent
$w_{\rm DE}=P/\rho$ while maintaining an exactly FLRW background.

This example is intended to demonstrate a possible background history rather than establish a fully viable model. Establishing viability additionally
requires the absence of ghost and gradient instabilities and consistency with
observational bounds on preferred-direction effects. 
The background evolution and perturbative signatures are controlled by the same fields and coefficients subject to the on-shell constraint and must
therefore be examined within a unified analysis.

\end{document}